\documentclass[summary]{ursi}
\usepackage{cite}
\usepackage{amsmath,amssymb,amsfonts}
\usepackage{amsmath}
\usepackage{amsthm}

\usepackage{graphicx}
\usepackage{textcomp}
\usepackage{xcolor}
\usepackage{graphicx}
\usepackage{float}
\usepackage{subfigure}
\usepackage{amsmath}
\usepackage{amsfonts,amssymb}
\usepackage{mathrsfs}
\usepackage{mathtools}
\usepackage{algpseudocode}
\usepackage{bm}
\usepackage{multirow}
\usepackage{array}
\usepackage{amssymb}
\usepackage{amsmath}
\usepackage{cite}
\usepackage{url}
\usepackage{xcolor}
\usepackage{cite,graphicx,amsmath,amssymb}
\usepackage{subfigure}
\usepackage{fancyhdr}
\usepackage{mdwmath}
\usepackage{mdwtab}
\usepackage{caption}
\usepackage{amsthm}
\usepackage{setspace}
\usepackage{bm}
\usepackage{algorithm}
\usepackage{algpseudocode}
\usepackage{mathtools}
\usepackage{dsfont}
\usepackage{bbm}
\usepackage{multicol}
\usepackage{framed}
\newtheorem{remark}{Remark}
\newtheorem{theorem}{Theorem}

\newtheorem{lemma}{Lemma}

\newtheorem{corollary}{Corollary}

\makeatletter
\newcommand{\biggg}{\bBigg@{3}}
\newcommand{\Biggg}{\bBigg@{3.5}}
\makeatother

\title{A Concise Tutorial for Analyzing Electromagnetic Degrees of Freedom for Continuous-Aperture Array (CAPA) Systems}

\author{Chongjun Ouyang\affref{ref1}, Boqun Zhao\affref{ref2}, Xingqi Zhang\affref{ref2}, and Yuanwei Liu\affref{ref3}}

\affiliation{%
  \aff{ref1}{School of Electronic Engineering and Computer Science, Queen Mary University of London, London, E1 4NS, U.K.}
  \aff{ref2}{Department of Electrical and Computer Engineering, University of Alberta, Edmonton AB, T6G 2R3, Canada}
  \aff{ref3}{Department of Electrical and Electronic Engineering, The University of Hong Kong, Hong Kong}
}

\begin{document}

\maketitle

\begin{abstract}
  A concise tutorial is provided for analysis of the spatial degrees of freedom (DoFs) in continuous-aperture array (CAPA)-based continuous electromagnetic (EM) channels. First, a simplified spatial model is introduced using the Fresnel approximation. By leveraging this model and Landau's theorem, a closed-form expression for the spatial DoFs is derived. The results show that the number of DoFs is proportional to the transmit and receive aperture sizes and inversely proportional to the propagation distance. Numerical results are presented to illustrate the properties of EM DoFs in CAPA-based channels.
\end{abstract}
\section{Introduction}
A continuous-aperture array (CAPA) is a promising multiple-antenna technology, essentially functioning as a spatially continuous electromagnetic (EM) surface \cite{ouyang2024diversity}. A CAPA operates as a single, large-aperture antenna with a continuous current distribution, comprising a (virtually) infinite number of radiating elements integrated with electronic circuits and driven by a limited number of radio-frequency (RF) chains. Unlike conventional spatially discrete arrays (SPDAs), a CAPA fully exploits the entire aperture surface and enables precise control over the current distribution \cite{ouyang2024diversity}. This capability significantly enhances spatial degrees of freedom (DoFs), as demonstrated in various research endeavors; see \cite{liu2024near} for more details.

However, the continuous nature of EM field interactions in CAPAs necessitates a fundamental shift in system modeling \cite{liu2024near,ouyang2024primer}. One of the most critical performance metrics requiring revision is the spatial DoFs. In SPDA systems, the spatial channel is represented as a discrete matrix, which can be decomposed into parallel, non-interfering sub-channels via singular value decomposition (SVD). These sub-channels enable optimal power allocation through water-filling to achieve channel capacity. The number of these sub-channels defines the system’s spatial DoFs, which determine the dominant communication modes available in the spatial channel.

In contrast, characterizing the spatial DoFs of CAPAs is significantly more challenging due to their continuous spatial responses, which are modeled as an integral operator rather than a discrete matrix. In realistic propagation environments, the spatial response operator (or radiation operator) can be expanded using a complete orthogonal basis set for both transmit currents and received fields, utilizing Hilbert-Schmidt decomposition or SVD \cite{migliore2008electromagnetics,migliore2018horse,pizzo2022spatial}. However, this approach is computationally demanding and does not yield explicit insights into spatial DoFs, as a closed-form expression for DoFs remains unavailable.

This paper aims to provide a concise tutorial for newcomers to analyze the number of DoFs in CAPA-based continuous EM channels. Our focus is on a point-to-point CAPA-based system. The central idea of this work is to use Landau's theorem to characterize the singular values (SVs) of the continuous spatial response operator. The main derivations in this tutorial are based on the original work presented in \cite{pizzo2022landau}, which is one of the first to utilize Landau's theorem to explore the number of DoFs between two CAPAs. While building on this foundation, we aim to make the derivations more accessible and easier to understand for newcomers in the field. In the interest of respecting the originality of prior work, we encourage readers to refer to \cite{pizzo2022landau} and related works to deepen their understanding of the application of Landau's theorem in CAPA-based systems, as well as the relationship between the spatial domain and the spatial-frequency domain (i.e., the wavenumber domain) \cite{pizzo2022spatial}. Other examples that utilize a similar approach to \cite{pizzo2022landau} include \cite{Do2023Line-of-sight,Do2023Parabolic,Ruiz-Sicilia2023MIMO}. The structure of this work is summarized as follows: \romannumeral1) we present a continuous spatial model for CAPAs to facilitate the analysis of spatial DoFs; \romannumeral2) using the simplified model, we leverage Landau's theorem to characterize the SVs of the continuous spatial response operator; \romannumeral3) we derive a closed-form expression for the EM DoFs, demonstrating that they are determined by the aperture size, propagation distance, and array geometry. \romannumeral4) we validate the effectiveness of the derived results through numerical simulations.

\begin{figure}[!t]
 \centering
\setlength{\abovecaptionskip}{0pt}
\includegraphics[height=0.18\textwidth]{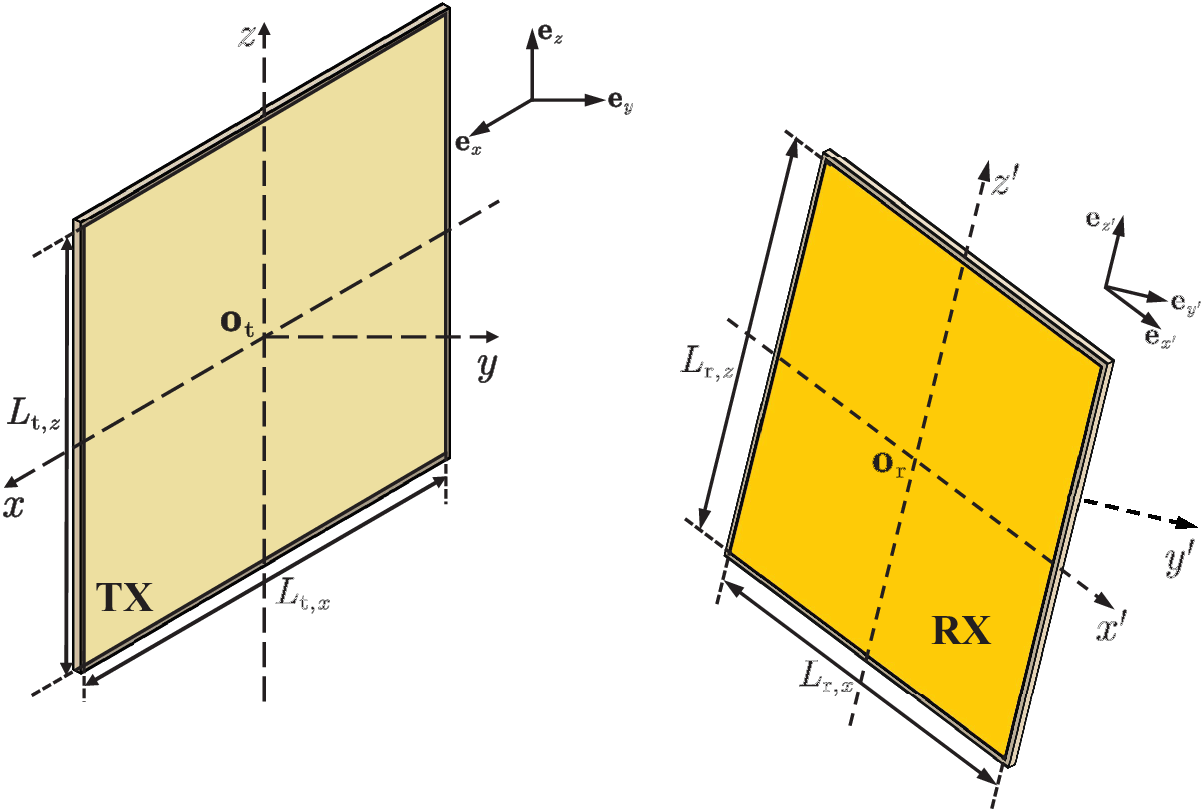}
\caption{Illustration of a CAPA-based channel.}
\label{Figure: Line-of-Sight Channel Model: System_Model}
\vspace{-10pt}
\end{figure}
\section{System Model}
Consider a point-to-point wireless communication system where both the transmitter (TX) and the receiver (RX) are equipped with a planar CAPA, as illustrated in {\figurename} {\ref{Figure: Line-of-Sight Channel Model: System_Model}}. The TX CAPA is positioned on the $x$-$z$ plane and centered at the origin ${\mathbf{o}}_{\rm{t}}=[0,0,0]^{T}$, with physical dimensions $L_{{\rm{t}},x}$ and $L_{{\rm{t}},z}$ along the $x$- and $z$-axes, respectively. The RX CAPA is centered at ${\mathbf{o}}_{\rm{r}}=[u_{x},u_{y},u_{z}]^{T}$ with a physical aperture of $L_{{\rm{r}},x}\times L_{{\rm{r}},z}$. As shown in {\figurename} {\ref{Figure: Line-of-Sight Channel Model: System_Model}}, the principal axes of the RX CAPA are denoted by $\mathbf{e}_{x'}\in{\mathbb{R}}^{3\times1}$ and $\mathbf{e}_{z'}\in{\mathbb{R}}^{3\times1}$, and its orientation is aligned along $\mathbf{e}_{y'}\in{\mathbb{R}}^{3\times1}$. Notably, ${\mathbf{E}}\triangleq [\mathbf{e}_{x'},\mathbf{e}_{y'},\mathbf{e}_{z'}]=\left[\begin{smallmatrix}e_{xx}&e_{xy}&e_{xz}\\
e_{yx}&e_{yy}&e_{yz}\\
e_{zx}&e_{zy}&e_{zz}\end{smallmatrix}\right]\in{\mathbb{R}}^{3\times3}$ forms an orthonormal basis in ${\mathbb{R}}^{3\times1}$, which satisfies ${\mathbf{E}}^{T}{\mathbf{E}}={\mathbf{E}}{\mathbf{E}}^{T}={\mathbf{I}}_3$. On this basis, another $x'y'z'$ Cartesian coordinate system can be constructed, with its origin at ${\mathbf{o}}_{\rm{r}}$ and its $x'$-, $y'$-, and $z'$-axes aligned with $\mathbf{e}_{x'}$, $\mathbf{e}_{y'}$, and $\mathbf{e}_{z'}$, respectively, as shown in {\figurename} {\ref{Figure: Line-of-Sight Channel Model: System_Model}}. The matrix ${\mathbf{E}}$ serves as a rotation matrix that transforms a point’s coordinates from the $x'y'z'$ system to the $xyz$ system. Let ${\mathbf{r}}=[r_x,r_y,r_z]^{T}$ represent a point's coordinates in the $x'y'z'$ system. Then, its coordinates in the $xyz$ system are given by ${\mathbf{o}}_{\rm{r}}+{\mathbf{E}}{\mathbf{r}}$. Therefore, the apertures of the TX and RX arrays are given by
\begin{subequations}
\begin{align}
{\mathcal{A}}_{\rm{t}}&=\{\left[x,0,z\right]^{T}|\delta\in[\begin{smallmatrix}-\frac{L_{{\rm{t}},\delta}}{2},\frac{L_{{\rm{t}},\delta}}{2}\end{smallmatrix}]~{\rm{for}}~\delta=x,z\},\\
{\mathcal{A}}_{\rm{r}}&=\{\left[x,0,z\right]^{T}|\delta\in[\begin{smallmatrix}-\frac{L_{{\rm{r}},\delta}}{2},\frac{L_{{\rm{r}},\delta}}{2}\end{smallmatrix}]~{\rm{for}}~\delta=x,z\}.
\end{align}
\end{subequations} 

The spatial channel response from ${\mathbf{t}}\in{\mathcal{A}}_{\rm{t}}$ to ${\mathbf{r}}\in{\mathcal{A}}_{\rm{r}}$ can be written as follows \cite{ouyang2024primer}: 
\begin{align}
h({\mathbf{r}},{\mathbf{t}})=\frac{-{\rm{j}}\eta_0k_0{\rm{e}}^{-{\rm{j}}k_0\lVert{\mathbf{o}}_{\rm{r}}+{\mathbf{E}}{\mathbf{r}}-{\mathbf{t}}\rVert}}
{4\pi\lVert{\mathbf{o}}_{\rm{r}}+{\mathbf{E}}{\mathbf{r}}-{\mathbf{t}}\rVert},
\end{align}
where $k_0=\frac{2\pi}{\lambda}$ is the wavenumber, with $\lambda$ denoting the wavelength, and $\eta$ represents the impedance of free space. Since the array apertures are generally negligible relative to the propagation distance $\lVert{\mathbf{o}}_{\rm{r}}+{\mathbf{E}}{\mathbf{r}}-{\mathbf{t}}\rVert$, the magnitude $\lvert h({\mathbf{r}},{\mathbf{t}})\rvert$ is approximately constant across ${\mathbf{r}}\in{{\mathcal{A}}_{\rm{r}}}$ and ${\mathbf{t}}\in{{\mathcal{A}}_{\rm{t}}}$ while $\lVert{\mathbf{o}}_{\rm{r}}+{\mathbf{E}}{\mathbf{r}}-{\mathbf{t}}\rVert\approx \lVert{\mathbf{o}}_{\rm{r}}-{\mathbf{o}}_{\rm{t}}\rVert$ such that only the phase variations need to be modeled. Thus, we have
\begin{equation}\label{dyadic Green's function_Standard_Scalar_USW}
h({\mathbf{r}},{\mathbf{t}})\approx
\frac{-{\rm{j}}\eta_0k_0{\rm{e}}^{-{\rm{j}}k_0\lVert{\mathbf{o}}_{\rm{r}}+{\mathbf{E}}{\mathbf{r}}-{\mathbf{t}}\rVert}}{4\pi D},
\end{equation}
where $D=\lVert{\mathbf{o}}_{\rm{r}}-{\mathbf{o}}_{\rm{t}}\rVert=\lVert{\mathbf{o}}_{\rm{r}}-[0,0,0]^{T}\rVert=\lVert{\mathbf{o}}_{\rm{r}}\rVert$. Using the Fresnel approximation, we obtain \cite{ouyang2024primer}
\begin{align}\label{Fresnel_Approximation}
\lVert({\mathbf{o}}_{\rm{r}}+{\mathbf{E}}{\mathbf{r}})-{\mathbf{t}}\rVert
&\approx D\left(1+\frac{{\mathbf{o}}_{\rm{r}}^{T}({\mathbf{E}}{\mathbf{r}}-{\mathbf{t}})}{D^2}+\frac{\lVert{\mathbf{E}}{\mathbf{r}}-{\mathbf{t}}\rVert^2}{2D^2}\right.\nonumber\\
&-\left.\frac{{\mathbf{o}}_{\rm{r}}^{T}({\mathbf{E}}{\mathbf{r}}-{\mathbf{t}})}{2D^3}\right).
\end{align} 
It follows that
\begin{equation}\label{Fresnel_approximation_used_DoF}
\begin{split}
h({\mathbf{r}},{\mathbf{t}})=\frac{-{\rm{j}}\eta_0k_0}{4\pi D}{\rm{e}}^{-{\rm{j}}k_0\left(D+\frac{{\mathbf{o}}_{\rm{r}}^{T}({\mathbf{E}}{\mathbf{r}}-{\mathbf{t}})}{D}+\frac{\lVert{\mathbf{E}}{\mathbf{r}}-{\mathbf{t}}\rVert^2}{2D}
-\frac{{\mathbf{o}}_{\rm{r}}^{T}({\mathbf{E}}{\mathbf{r}}-{\mathbf{t}})}{2D^2}\right)}.
\end{split}
\end{equation} 
\section{Analysis of the Number of EM DoFs}\label{Section: Analysis of the Number of EM DoFs}
We next characterize the EM DoFs provided by the continuous EM channel $h({\mathbf{r}},{\mathbf{t}})$. Through Hilbert-Schmidt decomposition or SVD, $h({\mathbf{r}},{\mathbf{t}})$ can be written as follows \cite{migliore2008electromagnetics,migliore2018horse,pizzo2022spatial}:
\begin{align}\label{CAPA_SU_SVD}
h({\mathbf{r}},{\mathbf{t}})=\sum\nolimits_{i=1}^{\infty}\phi_{i}({\mathbf{r}})\psi_{i}^{*}({\mathbf{t}})\sqrt{\lambda_{i}},
\end{align}
where $\sqrt{\lambda_{1}}\geq\sqrt{\lambda_{2}}\ldots\geq\sqrt{\lambda_{\infty}}\geq0$ are the SVs of $h({\mathbf{r}},{\mathbf{t}})$, and $\{\phi_{i}(\cdot)\}_{i=1}^{\infty}$ and $\{\psi_{i}(\cdot)\}_{i=1}^{\infty}$ are the singular functions. The sets $\{\psi_{i}(\cdot)\}_{i=1}^{\infty}$ and $\{\phi_{i}(\cdot)\}_{i=1}^{\infty}$ form orthonormal bases over ${\mathcal{A}}_{{\rm{t}}}$ and ${\mathcal{A}}_{{\rm{r}}}$, respectively. It holds that $\int_{{\mathcal{A}}_{\rm{r}}}\phi_{i}^{*}({\mathbf{r}})\phi_{i'}({\mathbf{r}}){\rm{d}}{\mathbf{r}}=\delta_{i,i'}$ and $\int_{{\mathcal{A}}_{\rm{t}}}\psi_{i}^{*}({\mathbf{t}})\psi_{i'}({\mathbf{t}}){\rm{d}}{\mathbf{t}}=\delta_{i,i'}$, where $\delta_{i,i'}$ is the Kronecker delta. Once \eqref{CAPA_SU_SVD} is determined, one can use the water-filling approach to compute CAPA capacity.

By removing the constant term $\frac{-{\rm{j}}\eta_0k_0}{4\pi D}{\rm{e}}^{-{\rm{j}}k_0D}$, we simplify $h({\mathbf{r}},{\mathbf{t}})$ as follows:
\begin{equation}\label{Fresnel_approximation_used_DoF_3ed}
\overline{h}({\mathbf{r}},{\mathbf{t}})={\rm{e}}^{-{\rm{j}}k_0\left(\frac{{\mathbf{o}}_{\rm{r}}^{T}({\mathbf{E}}{\mathbf{r}}-{\mathbf{t}})}{D}
+\frac{\lVert{\mathbf{E}}{\mathbf{r}}-{\mathbf{t}}\rVert^2}{2D}
-\frac{{\mathbf{o}}_{\rm{r}}^{T}({\mathbf{E}}{\mathbf{r}}-{\mathbf{t}})}{2D^2}\right)},
\end{equation}
whose SVs are given by $\{\frac{\eta_0k_0}{4\pi D}\sqrt{\lambda_i}\}_{i=0}^{\infty}$. Expanding and rearranging the terms in \eqref{Fresnel_approximation_used_DoF_3ed}, we have
\begin{equation}\label{Fresnel_approximation_used_DoF_4ed}
\overline{h}({\mathbf{r}},{\mathbf{t}})=\overline{\varphi}({\mathbf{E}}{\mathbf{r}}){\rm{e}}^{{\rm{j}}\frac{k_0}{D}{\mathbf{t}}^{T}{\mathbf{E}}{\mathbf{r}}}
\overline{\varphi}^{*}({\mathbf{t}}),
\end{equation}
where $\overline{\varphi}({\mathbf{x}})={\rm{e}}^{-{\rm{j}}k_0\left(\frac{{\mathbf{o}}_{\rm{r}}^{T}{\mathbf{x}}}{D}
+\frac{\lVert{\mathbf{x}}\rVert^2}{2D}-\frac{{\mathbf{o}}_{\rm{r}}^{T}{\mathbf{x}}}{2D^2}\right)}$. This kernel includes two separable quadratic phase shifts and a cross phase shift that depends on the TX and RX geometry. We further note that
\begin{align}
\overline{\varphi}({\mathbf{x}})\overline{\varphi}^{*}({\mathbf{x}})={\rm{e}}^{0}=1, 
\end{align}
which implies that $\overline{\varphi}({\mathbf{E}}{\mathbf{r}})$ and $\overline{\varphi}^{*}({\mathbf{t}})$ can be compensated using geometric knowledge at both ends of the link, and they do not affect the SVs. Therefore, $\overline{h}({\mathbf{r}},{\mathbf{t}})$ has the same SVs as ${\rm{e}}^{{\rm{j}}\frac{k_0}{D}{\mathbf{t}}^{T}{\mathbf{E}}{\mathbf{r}}'}$. By further noting that the $y$-component has no impact on the SVD of ${\rm{e}}^{{\rm{j}}\frac{k_0}{D}{\mathbf{t}}^{T}{\mathbf{E}}{\mathbf{r}}'}$, we omit the $y$-component in both ${\mathbf{r}}\in{{\mathcal{A}}_{\rm{r}}}$ and ${\mathbf{t}}\in{{\mathcal{A}}_{\rm{t}}}$, which simplifies ${\rm{e}}^{{\rm{j}}\frac{k_0}{D}{\mathbf{t}}^{T}{\mathbf{E}}{\mathbf{r}}}$ as follows:
\begin{align}
{\rm{e}}^{{\rm{j}}\frac{k_0}{D}{\mathbf{t}}^{T}{\mathbf{E}}{\mathbf{r}}}
={\rm{e}}^{{\rm{j}}\frac{k_0}{D}\left[t_x,t_z\right]{{\mathbf{E}}'}
\left[r_x ,r_z\right]^{T}},
\end{align}
where ${{\mathbf{E}}'}=\left[\begin{smallmatrix}e_{xx}&e_{xz}\\
e_{zx}&e_{zz}\end{smallmatrix}\right]\in{\mathbbmss{R}}^{2\times2}$. Furthermore, we define
\begin{subequations}
\begin{align}
{\mathcal{B}}_{\upsilon,\rm{t}}&\triangleq\{\left[x,z\right]^{T}|\delta\in[\begin{smallmatrix}-\frac{L_{{\rm{t}},\delta}\upsilon}{2},\frac{L_{{\rm{t}},\delta}\upsilon}{2}\end{smallmatrix}]~{\rm{for}}~\delta=x,z\},\\
{\mathcal{B}}_{\upsilon,\rm{r}}&\triangleq\{\left[x,z\right]^{T}|\delta\in[\begin{smallmatrix}-\frac{L_{{\rm{r}},\delta}\upsilon}{2},\frac{L_{{\rm{r}},\delta}\upsilon}{2}\end{smallmatrix}]~{\rm{for}}~\delta=x,z\}.
\end{align}
\end{subequations} 

Denote the SVD of ${\rm{e}}^{{\rm{j}}\frac{k_0}{D}{\mathbf{t}}^{T}{{\mathbf{E}}'}{\mathbf{r}}}$ as follows:
\begin{align}\label{To_User_Landau_Step1}
{\rm{e}}^{{\rm{j}}\frac{k_0}{D}{\mathbf{t}}^{T}{{\mathbf{E}}'}{\mathbf{r}}}=\sum\nolimits_{i=1}^{\infty}\hat{\phi}_{i}({\mathbf{r}})\hat{\psi}_{i}^{*}({\mathbf{t}})\sqrt{\sigma_i},
\end{align}
where $\sqrt{\sigma_{1}}\geq\sqrt{\sigma_{2}}\ldots\geq\sqrt{\sigma_{\infty}}\geq0$ are the SVs, and $\{\hat{\phi}_{i}(\cdot)\}_{i=1}^{\infty}$ and $\{\hat{\psi}_{i}(\cdot)\}_{i=1}^{\infty}$ are the associated singular functions defined on ${\mathbf{r}}=[r_x,r_z]^{T}\in{{\mathcal{B}}_{1,\rm{r}}}$ and ${\mathbf{t}}=[t_x,t_z]^{T}\in{{\mathcal{B}}_{1,\rm{t}}}$, respectively. For convenience, we apply a scaling transformation to the TX and RX apertures by a factor of $\upsilon_0=\sqrt{{k_0}/{D}}$ as follows:
\begin{subequations}
\begin{align}
{\mathbf{t}}&=\left[t_x,t_z\right]^{T}\Rightarrow{\mathbf{t}}=\left[t_x\upsilon_0,t_z\upsilon_0\right]^{T}\in{\mathcal{B}}_{\upsilon_0,\rm{t}},\\
{\mathbf{r}}&=\left[r_x,r_z\right]^{T}\Rightarrow{\mathbf{r}}=\left[r_x\upsilon_0,r_z\upsilon_0\right]^{T}\in{\mathcal{B}}_{\upsilon_0,\rm{r}}.
\end{align}
\end{subequations}
The scaled aperture regions satisfy
\begin{align}
m({\mathcal{B}}_{\upsilon_0,\rm{t}})=\upsilon_0^2m({\mathcal{B}}_{1,\rm{t}}),~m({\mathcal{B}}_{\upsilon_0,\rm{r}})=\upsilon_0^2m({\mathcal{B}}_{1,\rm{r}}),
\end{align}
where $m(\cdot)$ denotes the Lebesgue measure. This transformation simplifies ${\rm{e}}^{{\rm{j}}\frac{k_0}{D}{\mathbf{t}}^{T}{{\mathbf{E}}'}{\mathbf{r}}}$ to the follows:
\begin{align}\label{Transformed_CAPA_SU_SVD}
{\rm{e}}^{{\rm{j}}\frac{k_0}{D}{\mathbf{t}}^{T}{{\mathbf{E}}'}{\mathbf{r}}}\Rightarrow{\rm{e}}^{{\rm{j}}{\mathbf{t}}^{T}{{\mathbf{E}}'}{\mathbf{r}}}=\sum\nolimits_{i=1}^{\infty}\hat{\phi}_{i}({\mathbf{r}}/\upsilon_0)\hat{\psi}_{i}^{*}({\mathbf{t}}/\upsilon_0)\sqrt{\sigma_i},
\end{align}
where $\{\frac{1}{\upsilon_0}\hat{\phi}_{i}(\frac{{\mathbf{r}}}{\upsilon_0})\}_{i=1}^{\infty}$ and $\{\frac{1}{\upsilon_0}\hat{\psi}_{i}(\frac{\mathbf{t}}{\upsilon_0})\}_{i=1}^{\infty}$ form orthonormal bases over ${\mathbf{r}}\in{\mathcal{B}}_{\upsilon_0,\rm{r}}$ and $\mathbf{t}\in{\mathcal{B}}_{\upsilon_0,\rm{t}}$, respectively. Thus, the SVs of ${\rm{e}}^{{\rm{j}}{\mathbf{t}}^{T}{{\mathbf{E}}'}{\mathbf{r}}}$ on ${\mathcal{B}}_{\upsilon_0,\rm{r}}\times{\mathcal{B}}_{\upsilon_0,\rm{t}}$ are given by $\{\sqrt{\sigma_i}\upsilon_0^2\}_{i=1}^{\infty}$.

By \eqref{Transformed_CAPA_SU_SVD}, the autocorrelation at the TX side is given by
\begin{subequations}
\begin{align}
R_{\rm{t}}({\mathbf{t}},{\mathbf{t}}')&\triangleq\int_{{\mathcal{B}}_{\upsilon_0,\rm{r}}} {\rm{e}}^{{\rm{j}}{\mathbf{t}}^{T}{\mathbf{E}}'{\mathbf{r}}}{\rm{e}}^{-{\rm{j}}{\mathbf{t}'}^{T}{\mathbf{E}}'{\mathbf{r}}}{\rm{d}}{\mathbf{r}}\\
&=\sum\nolimits_{i,i'=1}^{\infty}\sqrt{\sigma_{i}\sigma_{i'}}\hat{\psi}_{i}^{*}({\mathbf{t}}/\upsilon_0)\hat{\psi}_{i'}({\mathbf{t}}'/\upsilon_0)\nonumber\\
&\times\int_{{\mathcal{B}}_{\upsilon_0,\rm{r}}}\hat{\phi}_{i}({\mathbf{r}}/\upsilon_0)\hat{\phi}_{i'}^{*}({\mathbf{r}}/\upsilon_0){\rm{d}}{\mathbf{r}}.\label{CAPA_SU_Correlation_RX_Step1}
\end{align}
\end{subequations}
Using the orthogonality of $\{\frac{1}{\upsilon_0}\hat{\phi}_{i}(\frac{{\mathbf{r}}}{\upsilon_0})\}_{i=1}^{\infty}$, we obtain
\begin{equation}\label{CAPA_SU_Correlation_TX_Simplified}
\begin{split}
R_{\rm{t}}({\mathbf{t}},{\mathbf{t}}')=\sum\nolimits_{i=1}^{\infty}{\sigma_{i}}\upsilon_0^2\hat{\psi}_{i}^{*}({\mathbf{t}}/\upsilon_0)\hat{\psi}_{i}({\mathbf{t}}'/\upsilon_0).
\end{split}
\end{equation}
On the other hand, the autocorrelation can also be given by
\begin{align}
R_{\rm{t}}({\mathbf{t}},{\mathbf{t}}')=\int_{{\mathcal{B}}_{\upsilon_0,\rm{r}}} {\rm{e}}^{{\rm{j}}\left[t_x-t_x',t_z-t_z'\right]{{\mathbf{E}}'}
\left[r_x,r_z\right]^{T}}{\rm{d}}r_x{\rm{d}}r_z.
\end{align}
Applying a coordinate transformation $\left[r_x,r_z\right]^{T}=\left[\begin{smallmatrix}e_{xx}&e_{xz}\\
e_{zx}&e_{zz}\end{smallmatrix}\right]\left[r_x,r_z\right]^{T}$, we obtain
\begin{align}\label{CAPA_SU_Correlation_RX_Landau_Step1}
R_{\rm{t}}({\mathbf{t}},{\mathbf{t}}')
=\int_{{\mathcal{B}}_{\upsilon_0,\rm{r}}'} {\rm{e}}^{{\rm{j}}\left[t_x-t_x',t_z-t_z'\right]
\left[\begin{smallmatrix}r_x\\r_z\end{smallmatrix}\right]}\lvert\det({{\mathbf{E}}'}^{-1})\rvert{\rm{d}}r_x{\rm{d}}r_z,
\end{align}
where $\lvert\det({{\mathbf{E}}'}^{-1})\rvert$ is the Jacobian determinant of the transformation. The transformed region ${\mathcal{B}}_{\upsilon_0,\rm{r}}'$ is given by
\begin{align}
{\mathcal{B}}_{\upsilon_0,\rm{r}}'=\left\{\left[\begin{smallmatrix}x\\z\end{smallmatrix}\right]={{\mathbf{E}}'}
\left[\begin{smallmatrix}x\\z\end{smallmatrix}\right]\left|\left[\begin{smallmatrix}x\\z\end{smallmatrix}\right]\in{\mathcal{B}}_{\upsilon_0,\rm{r}}\right.\right\}.
\end{align}
The aperture size covered by ${\mathcal{B}}_{\upsilon_0,\rm{r}}'$ satisfies
\begin{subequations}
\begin{align}
m({\mathcal{B}}_{\upsilon_0,\rm{r}})&=\int_{{\mathcal{B}}_{\upsilon_0,\rm{r}}}{\rm{d}}r_x{\rm{d}}r_z
=\int_{{\mathcal{B}}_{\upsilon_0,\rm{r}}'}\lvert\det({{\mathbf{E}}'}^{-1})\rvert{\rm{d}}r_x{\rm{d}}r_z\\
&={m({\mathcal{B}}_{\upsilon_0,\rm{r}}')}{\lvert\det({{\mathbf{E}}'})\rvert^{-1}}.
\end{align}
\end{subequations}

By removing $\lvert\det({{\mathbf{E}}'}^{-1})\rvert$ and multiplying by $\frac{1}{(2\pi)^2}$, \eqref{CAPA_SU_Correlation_RX_Landau_Step1} simplifies to the following:
\begin{align}\label{CAPA_SU_Correlation_RX_Landau_Step2}
\hat{R}_{\rm{t}}({\mathbf{t}},{\mathbf{t}}')
=\frac{1}{(2\pi)^2}\int_{{\mathcal{B}}_{\upsilon_0,\rm{r}}'} {\rm{e}}^{{\rm{j}}((t_x-t_x')r_x+(t_z-t_z')r_z)}{\rm{d}}r_x{\rm{d}}r_z.
\end{align}
For clarity, let $\varepsilon_1\geq \varepsilon_2\ldots\geq \varepsilon_{\infty}\geq0$ denote the eigenvalues of $\hat{R}_{\rm{t}}({\mathbf{t}},{\mathbf{t}}')$. By combining the results in \eqref{To_User_Landau_Step1} and \eqref{CAPA_SU_Correlation_TX_Simplified} with \eqref{CAPA_SU_Correlation_RX_Landau_Step2}, we conclude that $\{\sqrt{\lambda_i}\}_{i=1}^{\infty}$ satisfy
\begin{subequations}\label{EDoF_Planar_LoS_Operator_Equal_Result0}
\begin{align}
{{\lambda}_i}&=\varepsilon_i\lvert\det({{\mathbf{E}}'}^{-1})\rvert(2\pi)^2\frac{1}{\upsilon_0^4}\left(\frac{\eta_0k_0}{4\pi D}\right)^2\\
&=\frac{\varepsilon_i}{\lvert\det({{\mathbf{E}}'})\rvert}\left(\frac{\eta_0}{2 }\right)^2,~i=1,\ldots,\infty.
\end{align}
\end{subequations}
To further explore the behavior of the eigenvalues, consider an arbitrary square-integrable function $f_{{\rm{t}}}({\mathbf{t}})$ defined over ${\mathbf{t}}\in{\mathcal{B}}_{\upsilon_0,\rm{t}}$. We express $\overline{f}_{{\rm{t}}}({\mathbf{t}})\triangleq\int_{{\mathcal{B}}_{\upsilon_0,\rm{t}}}\hat{R}_{\rm{t}}({\mathbf{t}},{\mathbf{t}}')f_{{\rm{t}}}({\mathbf{t}}'){\rm{d}}{\mathbf{t}}'$ for ${\mathbf{t}}=[t_x,t_z]^{T}\in{\mathcal{B}}_{\upsilon_0,\rm{t}}$ as follows:
\begin{align}
\overline{f}_{{\rm{t}}}({\mathbf{t}})={\mathbbmss{1}}_{{\mathcal{B}}_{\upsilon_0,\rm{t}}}({\mathbf{t}})\int_{{\mathcal{B}}_{\upsilon_0,\rm{t}}}
\hat{R}_{\rm{t}}({\mathbf{t}},{\mathbf{t}}')
{\mathbbmss{1}}_{{\mathcal{B}}_{\upsilon_0,\rm{t}}}({\mathbf{t}}')f_{{\rm{t}}}({\mathbf{t}}'){\rm{d}}{\mathbf{t}}',
\end{align}
where $\hat{R}_{\rm{t}}({\mathbf{t}},{\mathbf{t}}')$ given in \eqref{CAPA_SU_Correlation_RX_Landau_Step2} represents the inverse Fourier transform of ${\mathbbmss{1}}_{{\mathcal{B}}_{\upsilon_0,\rm{r}}'}([r_x,r_z]^{T})$. In other words, the Fourier transform of $\hat{R}_{\rm{t}}({\mathbf{t}},{\mathbf{t}}')$ serves as an ideal filter over ${\mathcal{B}}_{\upsilon_0,\rm{r}}'$. 

Based on this setup, the properties of the eigenvalues $\{\varepsilon_{i}\}_{i=1}^{\infty}$ can be characterized using \emph{Landau's eigenvalue theorem}, which states \cite{landau1975szego,franceschetti2015landau}:
\begin{align}\label{EDoF_Planar_LoS_Operator_Equal_Result1}
1\geq \varepsilon_{1}\geq\varepsilon_{2}\ldots\geq\varepsilon_{\infty}\geq0
\end{align}
with $\{\varepsilon_{i}\}_{i=1}^{\infty}$ being functions of
\begin{align}\label{EDoF_Planar_LoS_Operator_Equal_Result_Important1}
{\mathsf{DOF}}=\frac{m({\mathcal{B}}_{\upsilon_0,\rm{r}}')m({\mathcal{B}}_{\upsilon_0,\rm{t}})}{(2\pi)^2}
=\frac{L_{{\rm{t}},x}L_{{\rm{t}},z}L_{{\rm{r}},x}L_{{\rm{r}},z}}{(\lambda D)^2\lvert\det({{\mathbf{E}}'})\rvert^{-1}}.
\end{align}
As $L_{{\rm{t}},x},L_{{\rm{t}},z},L_{{\rm{r}},x},L_{{\rm{r}},z}\rightarrow\infty$ or ${\mathsf{DOF}}\rightarrow\infty$, the eigenvalues $\{\varepsilon_{i}\}_{i=1}^{\infty}$ \emph{polarize} asymptotically \cite{landau1975szego,franceschetti2015landau}: for $\epsilon>0$, 
\begin{equation} \label{EDoF_Planar_LoS_Operator_Equal_Result2}
\begin{split}
&\lvert\{i:\varepsilon_{i}>\epsilon\}\rvert={\mathsf{DOF}}+o({\mathsf{DOF}}),
\end{split}
\end{equation}
as $L_{{\rm{t}},x},L_{{\rm{t}},z},L_{{\rm{r}},x},L_{{\rm{r}},z}\rightarrow\infty$ or ${\mathsf{DOF}}\rightarrow\infty$.
\vspace{-5pt}
\begin{remark}\label{EDoF_Planar_LoS_Operator_Equal_Result3}
The results in \eqref{EDoF_Planar_LoS_Operator_Equal_Result1} and \eqref{EDoF_Planar_LoS_Operator_Equal_Result2} imply that as the array aperture sizes $(L_{{\rm{t}},x},L_{{\rm{t}},z},L_{{\rm{r}},x},L_{{\rm{r}},z})$ increases, the leading ${\mathsf{DOF}}$ eigenvalues $\{\varepsilon_{i}\}_{i=1}^{{\mathsf{DOF}}}$ are close to one, while the remaining eigenvalues approach zero. This behavior resembles the step-like polarization pattern in \cite[{\figurename} 2]{franceschetti2015landau}.
\end{remark}
\vspace{-5pt}
Building on the findings in Remark \ref{EDoF_Planar_LoS_Operator_Equal_Result3} and \eqref{EDoF_Planar_LoS_Operator_Equal_Result0}, the SVs $\sqrt{\lambda_i}$, derived from the eigenvalues $\varepsilon_{i}$, exhibit a step-like behavior. For small values of $i$, these SVs remain approximately equal to $\frac{1}{\sqrt{\lvert\det({{\mathbf{E}}'})\rvert}}\frac{\eta_0}{2 }$ until reaching a critical index $i={\mathsf{DOF}}$, after which they rapidly decay. This step-like pattern becomes more pronounced as the array aperture sizes increase. Given that CAPAs are typically electromagnetically large arrays, satisfying $L_{{\rm{t}},x},L_{{\rm{t}},z},L_{{\rm{r}},x},L_{{\rm{r}},z}\gg \lambda$, the SVs of $h({\mathbf{r}},{\mathbf{t}})$ are dominated by the first ${\mathsf{DOF}}$ terms. This effective number of dominant SVs is commonly referred to as the effective DoFs (EDoFs), representing the number of primary communication sub-channels available for reliable transmission.

Recalling that the matrix ${\mathbf{E}}'$ in \eqref{EDoF_Planar_LoS_Operator_Equal_Result_Important1} is a submatrix of the rotation matrix ${\mathbf{E}}$, which can be expressed as follows: \cite{arfken2011mathematical}:
\begin{equation}\nonumber
\begin{split}
&{\mathbf{E}}^{T}=\\&\left[\begin{smallmatrix}
\cos{\alpha}\cos{\beta}&\cos{\alpha}\sin{\beta}\sin{\gamma}-\sin{\alpha}\cos{\gamma}&\cos{\alpha}\sin{\beta}\cos{\gamma}+\sin{\alpha}\sin{\gamma}\\
\sin{\alpha}\cos{\beta}&\sin{\alpha}\sin{\beta}\sin{\gamma}+\cos{\alpha}\cos{\gamma}&\sin{\alpha}\sin{\beta}\cos{\gamma}-\cos{\alpha}\sin{\gamma}\\
-\sin{\beta}&\cos{\beta}\sin{\gamma}&\cos{\beta}\cos{\gamma}
\end{smallmatrix}\right],
\end{split}
\end{equation}
where $\alpha$, $\beta$, and $\gamma$ represent the angles of counterclockwise rotation around the $z$-, $y$-, and $x$-axes, respectively, when viewed from the positive direction toward the origin. On this basis, we have
\begin{align}
\lvert\det({{\mathbf{E}}'})\rvert&=\left\lvert\det\left(\left[\begin{smallmatrix}\cos{\alpha}\cos{\beta}&-\sin{\beta}\\
\cos{\alpha}\sin{\beta}\cos{\gamma}+\sin{\alpha}\sin{\gamma}&\cos{\beta}\cos{\gamma}\end{smallmatrix}\right]\right)\right\rvert\nonumber\\
&=\lvert\cos{\alpha}\cos{\gamma}+\sin{\alpha}\sin{\gamma}\sin{\beta}\rvert.
\end{align}
It follows that
\begin{subequations}
\begin{align}
\lvert\det({{\mathbf{E}}'})\rvert&\leq\sqrt{\cos^2{\gamma}+\sin^2{\gamma}\sin^2{\beta}}
\leq \sqrt{\cos^2{\gamma}+\sin^2{\gamma}}=1.\nonumber
\end{align}
\end{subequations}
Furthermore, one configuration that achieves $\lvert\det({{\mathbf{E}}'})\rvert=1$ is when $\alpha=\gamma=0$, which means that the RX CAPA is parallel with the TX CAPA. These arguments imply that when the TX and RX CAPAs are not parallel in a LoS propagation environment, the number of EDoFs is reduced.

\begin{figure}[!t]
\centering
    \subfigbottomskip=0pt
	\subfigcapskip=-5pt
\setlength{\abovecaptionskip}{0pt}
    \subfigure[Parallel. $\alpha=\gamma=0$.]
    {
        \includegraphics[height=0.13\textwidth]{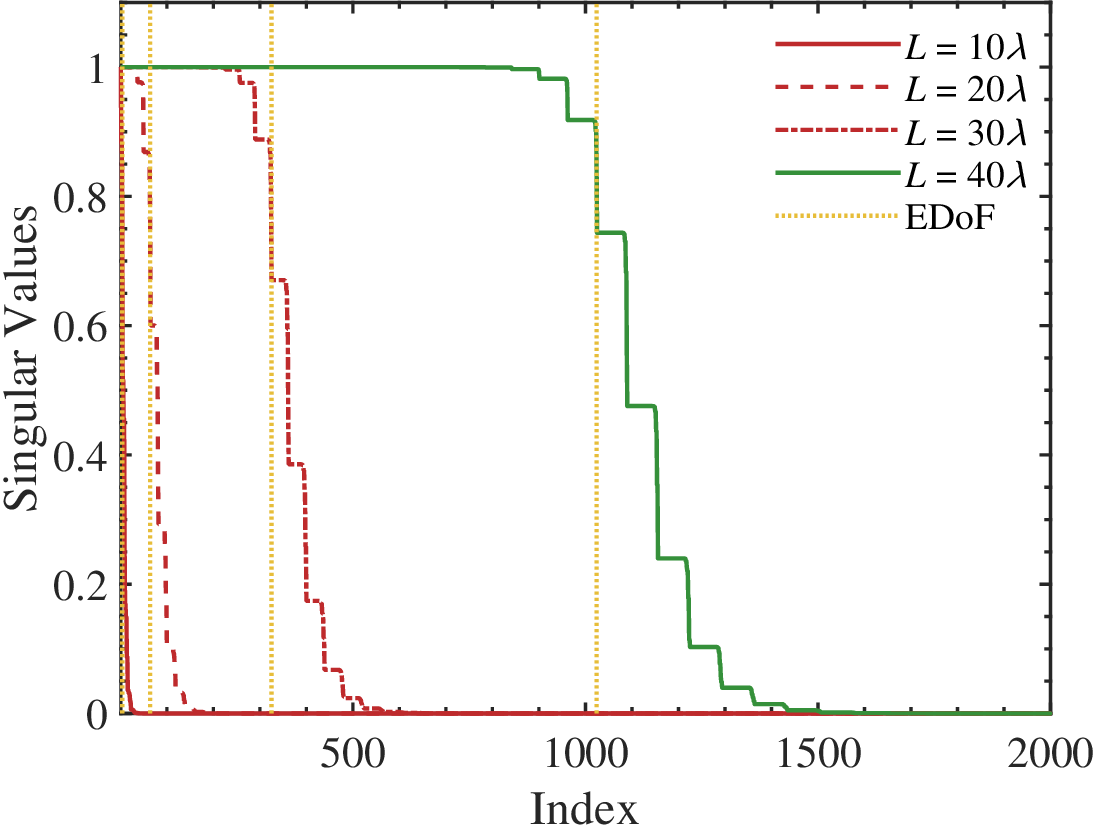}
	   \label{Figure: SVD_Parallel}	
    }
   \subfigure[Non-parallel. $\alpha=\gamma=\frac{\pi}{4}$.]
    {
        \includegraphics[height=0.13\textwidth]{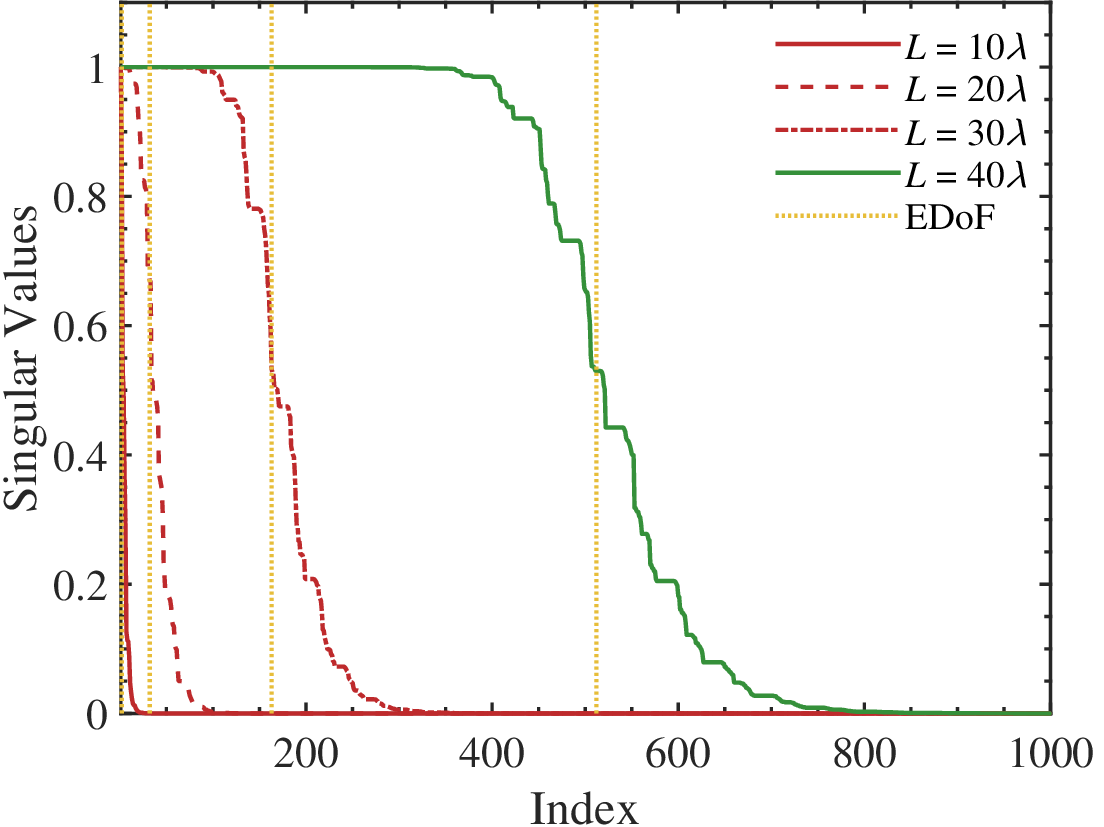}
	   \label{Figure: SVD_Non}	
    }
\caption{Singular values of the spatial response.}
\label{Figure: SVD}
\vspace{-10pt}
\end{figure}

\section{Numerical Results}
Numerical results are provided to validate the derived analytical findings. The following parameter setup is applied unless specified otherwise: the propagation distance is set as $D=50\lambda$, the carrier frequency is given by $f_{\rm{c}}=2.4$ GHz, $L_{{\rm{t}},x}=L_{{\rm{t}},z}=L_{{\rm{r}},x}=L_{{\rm{r}},z}=L$, ${\mathbf{o}}_{\rm{r}}=[0,D,0]^{T}$, and $\beta=0$. The SVs of the continuous operator are computed using the method detailed in \cite{atkinson2007solving}, and all presented SVs are normalized by their maximum value.

{\figurename} {\ref{Figure: SVD}} presents the ordered SVs of the spatial response operator $h({\mathbf{r}},{\mathbf{t}})$ for two scenarios: (a) when the TX CAPA is parallel to the RX CAPA (as per {\figurename} {\ref{Figure: SVD_Parallel}}), and (b) when the TX CAPA is not parallel to the RX CAPA (as per {\figurename} {\ref{Figure: SVD_Non}}). From {\figurename} {\ref{Figure: SVD}}, it is evident that in both cases, the SVs exhibit a step-like decay pattern. More specifically, when the SV index is smaller than the EDoF, the SVs decay slowly and remain close to one. When the SV index exceeds the EDoF, the SVs rapidly decay toward zero. These observations validate the theoretical derivations presented in Section \ref{Section: Analysis of the Number of EM DoFs}. Comparing {\figurename} {\ref{Figure: SVD_Parallel}} and {\figurename} {\ref{Figure: SVD_Non}}, we observe that parallel CAPA deployment results in a higher number of EDoFs than the non-parallel case. Furthermore, increasing the aperture size leads to a larger number of EDoFs, further enhancing the spatial resolution and communication capacity.

\begin{figure}[!t]
 \centering
\setlength{\abovecaptionskip}{0pt}
\includegraphics[height=0.15\textwidth]{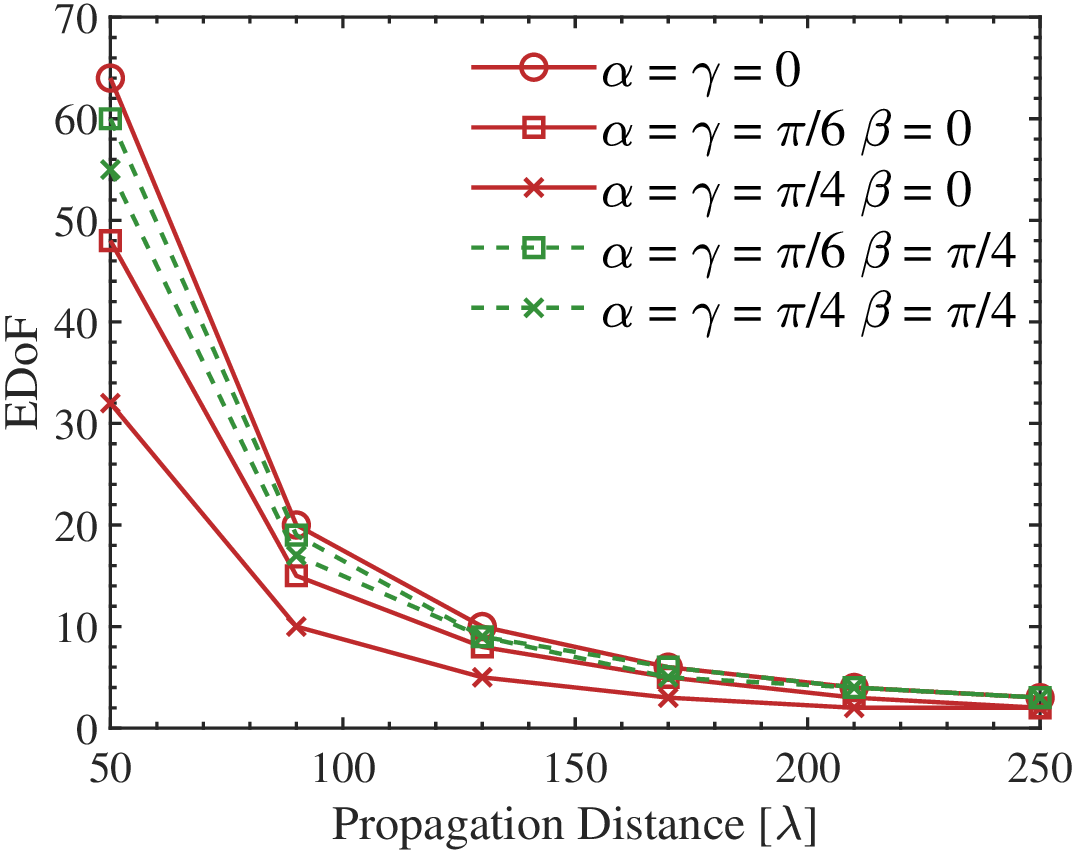}
\caption{Illustration of the EDoF.}
\label{Figure: EDoF}
\vspace{-10pt}
\end{figure}

{\figurename} {\ref{Figure: EDoF}} illustrates the relationship between the number of EDoFs and the propagation distance $D$. The results indicate that as $D$ increases, the number of EDoFs provided by CAPAs gradually decreases. This phenomenon occurs because, at larger propagation distances, the RX CAPA transitions into the far-field region of the TX CAPA. In this regime, EM wave propagation is dominated by plane-wave transmission, which inherently exhibits lower distance resolution and leads to a low-rank spatial channel.

\section{Conclusion}
We have presented a concise tutorial for analyzing the spatial DoFs in CAPA-based LoS channels. Our analysis demonstrates that the number of effective EDoFs is determined by the aperture size, propagation distance, and array geometry. Additionally, we have shown that the EDoFs are maximized when the TX and RX CAPAs are deployed in parallel configurations. We hope this concise tutorial will help newcomers in the field learn how to calculate the EDoFs. For a deeper understanding of this topic, we encourage interested readers to refer to \cite{pizzo2022landau} and \cite{franceschetti2015landau}.

\end{document}